# Nonlinear extension of the J-matrix method of scattering: A toy model

A. D. Alhaidari[(a)] and T. J. Taiwo[(b)]

[(a)] *Saudi Center for Theoretical Physics, P. O. Box 32741, Jeddah 21438, Saudi Arabia*
[(b)] *Physics Department, Untied Arab Emirate University, P.O. Box 15551, Al-Ain, United Arab Emirates*

**Abstract:** We introduce nonlinear extension of the J-matrix method of scattering. The formulation relies predominantly on the linearization of products of orthogonal polynomials. We present a toy model as an illustrative example and obtain the nonlinear scattering matrix.

## 1. Introduction

To study microscopic systems (e.g., atoms, molecules, nucleons, elementary particles, etc.), physicists conduct scattering experiments. Typically, in such an experiment, a beam of particles or energy (e.g., a beam of electrons or a laser ray) of well-defined properties is fired at the target under investigation. The output of the collision (the scattered matter or energy) is collected at the detectors; measured and studied. Comparing the properties of the incident and scattered objects gives valuable information about the target system. Such information is embedded in what is known as the scattering matrix (S-matrix) or phase shift that depends on the properties of the incident beam (e.g., its energy, angular momentum, spin, polarization, etc.) as well as the properties of the target. Usually, such scattering experiment is repeated for different values of incident beam properties. That is, for different energies, polarizations, angular momenta, etc. Scattering is divided into different categories depending on the outcome of the experiment. For example, if the nature of the scattered objects is identical to that of the incident objects then the scattering is termed elastic. Otherwise, the scattering process is inelastic (for example, the incident is a beam of protons whereas electrons are collected at the detectors). Another category of scattering experiments is the number of scattering channels that the process can go through. That happens when the properties of the scattered objects change dramatically from one range of values of the incident beam property to another range. For example, if the properties of the scattered objects change abruptly at certain critical values of the incident beam energy. In this work, we assume elastic and single channel scattering. Moreover, we characterize the scattering experiment by the solution of the wave equation (the nonrelativistic Schrödinger equation) for the incident particles as it goes through the process and we model the target system by a potential function. In that case, the scattering information could be extracted from the phase shift of the scattered particles relative to the incident. Additionally, we consider the potential in the wave equation to have a nonlinear component and use the tools of the J-matrix method of scattering in the solution of the problem. However, the J-matrix method was originally developed for linear interactions, thus we must start by extending it to handle nonlinear interactions.

In the atomic units $\hbar = m = 1$, Schrödinger's equation for the incident particles reads as follows

$$\mathrm{i}\partial_t \chi(t,x) = H\chi(t,x) = \left(H_0 + \mathcal{V}\right)\chi(t,x)\,. \tag{1}$$

For time-independent Hamiltonian, we can write $\chi(t,x) = e^{-\mathrm{i}Et}\psi_E(x)$ and choose $\psi_E(x) = \sum_n f_n(E)\phi_n(x)$, where $\{\phi_n(x)\}$ is a complete set of square integrable functions and $\{f_n(E)\}$ are expansion coefficients that contain all physical information about the system [1]. Moreover, the basis set is chosen such that the matrix representation of the reference Hamiltonian $H_0$ is tridiagonal and symmetric. That is, $\langle\phi_n|H_0|\phi_m\rangle = a_n\delta_{nm} + b_n\delta_{n,m-1} + b_{n-1}\delta_{n,m+1}$ where $\{a_n, b_n\}$ are constants such that $b_n \neq 0$ for all $n$. Additionally, we consider short-range potentials that consist of two parts; a linear component, $V$, and a nonlinear component, $W$. That is, $\mathcal{V} = V + W$ and for orthonormal basis, Eq. (1) is mapped into the corresponding projected energy space as follows

$$E\,f_n(E) = a_n\,f_n(E) + b_{n-1}\,f_{n-1}(E) + b_n\,f_{n+1}(E) + \sum\nolimits_{m=0}^{N-1} V_{nm}\,f_m(E) + \sum\nolimits_{m=0}^{M-1} W_{nm}(E)\,f_m(E)\,, \tag{2}$$

where $N$ and $M$ are large enough integers such that the matrix representation of the two parts of the short-range potential are accurate enough and $V_{nm} = \langle\phi_n|V|\phi_m\rangle$. In the linear theory where $W = 0$, Eq. (2) is solved as follows (see, for example, Part I in Ref. [2]). The matrix elements $\{V_{nm}\}_{n,m=0}^{N-1}$ are calculated numerically and added to $\left\{(H_0)_{nm}\right\}_{n,m=0}^{\infty}$ giving an infinite total Hamiltonian matrix that consists of an $N\times N$ finite block representing $H_0 + V$ and an infinite tridiagonal tail representing $H_0$. The infinite tail equation gives a three-term recursion relation, which is solved analytically in terms of orthogonal polynomials for $\{f_n(E)\}_{n=N-1}^{\infty}$. The remaining $N\times N$ block results in $N$ equations to be solved for $N$ unknowns. These unknowns are $\{f_n(E)\}_{n=0}^{N-2}$ and the phase shift $\delta(E)$. The latter gives the scattering matrix as $e^{2\mathrm{i}\delta(E)}$.

Now, to the nonlinear theory and for simplicity, we eliminate the linear part (i.e., take $V = 0$). Moreover, we consider a nonlinear *quartic* self-interacting potential, $|\chi(t,x)|^4$, such that we can write (without loss of generality) $W_{nm} = \sum_{ij}\alpha_{nm}^{ij} f_i f_j^*$, where $\{\alpha_{nm}^{ij}\}$ are constants such that $\left(\alpha_{nm}^{ij}\right)^* = \alpha_{mn}^{ji}$ and $\alpha_{nm}^{ij} = 0$ for $n,m \geq M$ (i.e., $W^\dagger = W$). In this case, the fundamental equation (2) becomes

$$E\,f_n(E) = a_n\,f_n(E) + b_{n-1}\,f_{n-1}(E) + b_n\,f_{n+1}(E) + \sum_{m=0}^{M-1}\sum_{i,j=0}^{K-1}\alpha_{nm}^{ij}\,f_i(E)\,f_j^*(E)\,f_m(E)\,, \tag{3}$$

for some positive integer $K$ such that $K \leq M$. Now, if we follow a procedure similar to the linear theory to find the scattering matrix then we face two problems:

(i) The nonlinear coupling of the expansion coefficients $\{f_n(E)\}$ in the last term of Eq. (3), and

(ii) The energy dependence of the matrix elements of the potential $\{W_{nm}(E)\}$.

In this work, we resolve the first and most serious problem by a linearization technique used for products of orthogonal polynomials [3]. The second problem will be resolved by a linear transformation of the sub-basis $\{\phi_n(x)\}_{n=0}^{M-1}$ into a special non-orthogonal finite basis. For simplicity in this work, we propose the following scenario: $\alpha_{nm}^{ij} = g\,\delta_{nm}\delta^{ij}$ making $W_{nm}(E) = g\,\delta_{nm}\sum_{i=0}^{K-1}[f_i(E)]^2$ where $g$ is a coupling parameter of inverse length dimension and we have assumed real coefficients, $f_i^*(E) = f_i(E)$. In Section 2, we formulate the nonlinear problem within this scenario and in Section 3 we obtain its solution for single channel elastic scattering.

## 2. Formulation of the problem

Let us consider a problem where the reference Hamiltonian is the kinetic energy operator in 3D with spherical symmetry. That is, we write

$$H_0 = -\frac{1}{2}\frac{d^2}{dr^2} + \frac{\ell(\ell+1)}{2r^2}, \tag{4}$$

where $\ell$ is the angular momentum quantum number. There are two independent solutions for the reference problem, which are written in terms of the Bessel functions of the first and second kind: the regular solution is $\psi_{reg}(r) = \sqrt{2kr}\,J_{\ell+\frac{1}{2}}(kr)$ and the irregular solution is $\psi_{irr}(r) = \sqrt{2kr}\,Y_{\ell+\frac{1}{2}}(kr)$, where $k = \sqrt{2E}$. On the other hand, the two corresponding J-matrix solutions of the reference problem, which are called the "sine-like, $S(r)$" and "cosine-like, $C(r)$", have already been established (see, for example, Section 2.3 in Ref. [4]). We choose a complete orthonormal basis (sometimes called the "oscillator basis") with the following elements

$$\phi_n(r) = \sqrt{2\,\Gamma(n+1)\big/\Gamma\left(n+\ell+\tfrac{3}{2}\right)}\left(\lambda r\right)^{\ell+1} e^{-\lambda^2 r^2/2} L_n^{\ell+\frac{1}{2}}(\lambda^2 r^2), \tag{5}$$

where $L_n^{\nu}(z)$ being the Laguerre polynomial and $\lambda$ is a scale parameter of inverse length dimension. In this basis set, the J-matrix reference solutions are written as

$$S(r) = \sum_{n=0}^{\infty} s_n(E)\phi_n(r), \qquad C(r) = \sum_{n=0}^{\infty} c_n(E)\phi_n(r). \tag{6}$$

The expansion coefficients $\{s_n\}$ and $\{c_n\}$ satisfy the following three-term recursion relation

$$\mu^2 P_n = \left(2n+\ell+\tfrac{3}{2}\right)P_n + \sqrt{n\left(n+\ell+\tfrac{1}{2}\right)}P_{n-1} + \sqrt{(n+1)\left(n+\ell+\tfrac{3}{2}\right)}P_{n+1}, \tag{7}$$

where $\mu = k/\lambda$ and $P_n$ stands for either $s_n$ or $c_n$ with the following initial relations ($n=0$):

$$\mu^2 s_0 = \left(\ell+\tfrac{3}{2}\right)s_0 + \sqrt{\ell+\tfrac{3}{2}}\,s_1, \tag{8a}$$

$$\mu^2 c_0 = \left(\ell+\tfrac{3}{2}\right)c_0 + \sqrt{\ell+\tfrac{3}{2}}\,c_1 - \frac{2}{\pi}\sqrt{\lambda^{-1}\Gamma\left(\ell+\tfrac{3}{2}\right)}\mu^{-\ell}\,e^{\mu^2/2}. \tag{8b}$$

The solutions of these recursion relations are [4]:

$$s_n(E) = 2(-1)^n \sqrt{\frac{\Gamma(n+1)}{\Gamma\left(n+\ell+\frac{3}{2}\right)}}\ \mu^{\ell+1} e^{-\mu^2/2} L_n^{\ell+\frac{1}{2}}(\mu^2), \tag{9a}$$

$$c_n(E) = \frac{2}{\pi}(-1)^n \Gamma\left(\ell+\tfrac{1}{2}\right)\sqrt{\frac{\Gamma(n+1)}{\Gamma(n+\ell+\frac{3}{2})}}\, \mu^{-\ell} e^{-\mu^2/2}\, {}_1F_1\left(-n-\ell-\tfrac{1}{2}, \tfrac{1}{2}-\ell, \mu^2\right). \tag{9b}$$

We should note that the sine-like solution $S(r)$ is identical to the regular reference solution $\psi_{reg}(r)$ everywhere whereas the cosine-like solution $C(r)$ is equal to the irregular reference solution $\psi_{irr}(r)$ only asymptotically (far away from the origin). The tridiagonal matrix representation of the reference Hamiltonian in the basis (5) reads as follows

$$\begin{aligned}\langle \phi_n | H_0 | \phi_m \rangle &= \frac{\lambda^2}{2}\left(2n+\ell+\tfrac{3}{2}\right)\delta_{nm} \\ &+ \frac{\lambda^2}{2}\sqrt{(n+1)\left(n+\ell+\tfrac{3}{2}\right)}\ \delta_{n,m-1} + \frac{\lambda^2}{2}\sqrt{n\left(n+\ell+\tfrac{1}{2}\right)}\ \delta_{n,m+1}\end{aligned} \tag{10}$$

giving the coefficients $\{a_n, b_n\}$ of Eq. (3). Due to the assumed short range of the nonlinear potential and renaming $M \mapsto N$, the wave equation (3) becomes the following matrix

$$\begin{pmatrix} \times & \times & \times & \times & \times & \times & & & & & & & \\ \times & \times & \times & \times & \times & \times & & & & & & & \\ \times & \times & \times & \times & \times & \times & & & & & & & \\ \times & \times & \times & \times & \times & \times & & & & & & & \\ \times & \times & \times & \times & \times & \times & & & & & & & \\ \times & \times & \times & \times & \times & \times & b_{N-1} & & & & & & \\ & & & & & b_{N-1} & a_N & b_N & & & & & \\ & & & & & & b_N & a_{N+1} & b_{N+1} & & & & \\ & & & & & & & b_{N+1} & a_{N+2} & b_{N+2} & & & \\ & & & & & & & & \times & \times & \times & & \\ & & & & & & & & & \times & \times & \times & \\ & & & & & & & & & & \times & \times & \times \\ & & & & & & & & & & & \times & \times \end{pmatrix} \begin{pmatrix} f_0 \\ f_1 \\ \times \\ \times \\ f_{N-2} \\ f_{N-1} \\ f_N \\ f_{N+1} \\ f_{N+2} \\ \times \\ \times \\ \times \\ \times \end{pmatrix} = E \begin{pmatrix} f_0 \\ f_1 \\ \times \\ \times \\ f_{N-2} \\ f_{N-1} \\ f_N \\ f_{N+1} \\ f_{N+2} \\ \times \\ \times \\ \times \\ \times \end{pmatrix} \tag{11}$$

The $N \times N$ block on the top left corner is the matrix representation of the total nonlinear Hamiltonian $H_0 + W$. The infinite sequence of equations resulting from row $N$ to infinity is just the three-term recursion relation (7) giving $f_n$ as linear combination of $s_n$ and $c_n$ for $n \geq N-1$. The boundary conditions make this combination read as follows [2,4]

$$f_n(E) = \left(c_n - \mathrm{i}s_n\right) - e^{2\mathrm{i}\delta(E)}\left(c_n + \mathrm{i}s_n\right) := F_n(E), \tag{12}$$

for $n = N-1, N, N+1, \ldots$ The remaining finite $N$ equations are

$$\left(\begin{array}{c} \\ \\ \\ H-E \\ \\ \\ \\ \end{array}\right)\begin{pmatrix} f_0 \\ f_1 \\ f_2 \\ \times \\ \times \\ f_{N-2} \\ F_{N-1} \end{pmatrix} = \begin{pmatrix} 0 \\ 0 \\ 0 \\ \times \\ \times \\ 0 \\ -b_{N-1}F_N \end{pmatrix} \tag{13}$$

where we have moved $E$ to the left side making the matrix on the left a finite $N\times N$ matrix representation of the wave operator $H-E=H_0+W-E$ in the orthonormal basis (5). The resulting $N$ equations will be solved for the $N$ unknowns $\left\{f_n(E)\right\}_{n=0}^{N-2}$ and $\delta(E)$. Let the inverse of the finite $N\times N$ matrix on the left (the finite Green's function of the problem) be named $G(E)$. Then multiplying both sides of Eq. (13) by $G(E)$, we obtain $N-1$ equations for $\left\{f_n\right\}_{n=0}^{N-2}$ that read

$$f_m(E)=-b_{N-1}G(E)_{m,N-1}F_N(E), \qquad m=0,1,...,N-2. \tag{14}$$

in addition to the following special relation $F_{N-1}(E)=-b_{N-1}G(E)_{N-1,N-1}F_N(E)$. Substituting $F_{N-1}(E)$ from Eq. (12) into this special relation gives the scattering matrix as follows [2]

$$\begin{aligned} e^{2\mathrm{i}\delta(E)} &= \frac{c_{N-1}(E)-\mathrm{i}s_{N-1}(E)+b_{N-1}G_{N-1,N-1}(E)\left[c_N(E)-\mathrm{i}s_N(E)\right]}{c_{N-1}(E)+\mathrm{i}s_{N-1}(E)+b_{N-1}G_{N-1,N-1}(E)\left[c_N(E)+\mathrm{i}s_N(E)\right]} \\ &= T_{N-1}(E)\frac{1+G_{N-1,N-1}(E)J_{N-1,N}(E)R_N^-(E)}{1+G_{N-1,N-1}(E)J_{N-1,N}(E)R_N^+(E)} \end{aligned} \tag{15}$$

where $T_n(E)=\dfrac{c_n(E)-\mathrm{i}s_n(E)}{c_n(E)+\mathrm{i}s_n(E)}$, $R_n^\pm(E)=\dfrac{c_n(E)\pm\mathrm{i}s_n(E)}{c_{n-1}(E)\pm\mathrm{i}s_{n-1}(E)}$, and $J_{n,m}(E)=\left(H_0\right)_{n.m}-E\delta_{n,m}$. In the following section, we obtain an explicit form for this S-matrix [i.e., write down the Green's function $G_{N-1,N-1}(E)$ explicitly] for a simple illustrative single channel elastic scattering example.

# 3. A toy model

We propose a simple nonlinear model by taking the expansion coefficients $\left\{f_n\right\}_{n=0}^{N-2}$ in the nonlinear wave equation (13) as weighted Laguerre polynomials in the energy. That is, we write

$$f_n(E):=\omega(E)\left[A_n\,L_n^\nu(\mu^2)\right], \tag{16}$$

where $\nu>-1$, $A_n=\sqrt{\Gamma(n+1)/\Gamma\left(n+\nu+1\right)}$ and $\omega(E)$ is some energy dependent weight function. If we take $\nu=\ell+\frac{1}{2}$ and $\omega(E)=2\mu^{\ell+1}e^{-\mu^2/2}$, then $f_n(E)=(-1)^n s_n(E)$. Using the ansatz (16), we can write

$$\sum_{m=0}^{N-1} W_{nm} f_m = g\, f_n \sum_{i=0}^{K-1} f_i^2 := g\, \omega^3(E) \sum_{i=0}^{K-1} \tilde{L}_n^\nu(\mu^2) \tilde{L}_i^\nu(\mu^2) \tilde{L}_i^\nu(\mu^2), \tag{17}$$

where $\tilde{L}_n^\nu(\mu^2)$ is the orthonormal version of the Laguerre polynomial. That is, the version that makes the orthogonality relation normalized (i.e., $\left\langle \tilde{L}_n^\nu(x) \middle| \tilde{L}_m^\nu(x) \right\rangle = \delta_{n,m}$) and the three-term recursion relation symmetric. Thus, $\tilde{L}_n^\nu(\mu^2) = A_n L_n^\nu(\mu^2)$. Using the linearization of products of orthogonal polynomials developed in Ref. [5], we can write the following third order linearization formula

$$\tilde{L}_i^\nu(z) \tilde{L}_i^\nu(z) \tilde{L}_n^\nu(z) = \sum_{m=0}^{n+2i} D_{i,n}^m(\nu) \tilde{L}_m^\nu(z), \tag{18}$$

where

$$D_{i,n}^m(\nu) = \left[ \tilde{L}_i^\nu(J) \tilde{L}_i^\nu(J) \right]_{n,m} \tag{19}$$

and $J$ is the following tridiagonal symmetric Jacobi matrix associated with the three-recursion relation of the orthonormal Laguerre polynomials, $z\tilde{L}_n^\nu(z) = \alpha_n \tilde{L}_n^\nu(z) + \beta_n \tilde{L}_{n+1}^\nu(z) + \beta_{n-1} \tilde{L}_{n-1}^\nu(z)$,

$$J = \begin{pmatrix} \alpha_0 & \beta_0 & & & \\ \beta_0 & \alpha_1 & \beta_1 & & \\ & \beta_1 & \alpha_2 & \beta_2 & \\ & & \times & \times & \times \\ & & & \times & \times \end{pmatrix}, \tag{20}$$

with $\alpha_n = 2n + \nu + 1$ and $\beta_n = -\sqrt{(n+1)(n+\nu+1)}$. Consequently, we can write (17) as

$$\sum_{m=0}^{N-1} W_{nm} f_m = g\omega^2(E) \sum_{i=0}^{K-1} \sum_{m=0}^{n+2i} D_{i,n}^m(\nu) f_m(E), \tag{21}$$

where we have rewritten $\tilde{L}_m^\nu(\mu^2)$ in Eq. (18) in terms of $f_m(E)$ using the definition (16). If we uphold the short-range assumption of the nonlinear coupling that we made at the end of Section 1, then the sum over the index *m* on the right side of Eq. (21) will terminate at $N-1$ and (21) becomes

$$\sum_{m=0}^{N-1} W_{nm} f_m = g\omega^2(E) \sum_{m=0}^{N-1} \Lambda_{n,m} f_m(E), \tag{22}$$

where $\Lambda_{n,m} = \sum_{i=0}^{K-1} D_{i,n}^m(\nu)$. In the Appendix, we prove that the matrix $\Lambda$ is positive definite (i.e., all of its eigenvalues are positive) as long as $\nu > -1$. Adding (22) to the reference wave equation, we obtain Eq. (11) with the $N \times N$ block on the top left corner being the matrix representation of the total nonlinear Hamiltonian $H_0 + W$, which is energy dependent due to the weight factor $\omega^2(E)$. Proceeding further, we obtain Eq. (13) that reads explicitly as follows

$$\left(\begin{array}{c} \\ \\ \\ H_0 + g\omega^2(E)\Lambda - EI \\ \\ \\ \\ \end{array}\right)\begin{pmatrix} f_0 \\ f_1 \\ f_2 \\ \times \\ \times \\ f_{N-2} \\ F_{N-1} \end{pmatrix} = \begin{pmatrix} 0 \\ 0 \\ 0 \\ \times \\ \times \\ 0 \\ -b_{N-1}F_N \end{pmatrix}, \qquad (23)$$

where $I$ is the $N \times N$ unit matrix and $H_0$ is the $N \times N$ tridiagonal symmetric matrix (10). Thus, we managed to resolve the first of the two problems pointed out below Eq. (3) by using the linearization technique developed in Ref. [5] and shown in Eq. (18). The task now is to solve Eq. (23) for $\{f_n(E)\}_{n=0}^{N-2}$ and $\delta(E)$ for all energies $E$. This is more difficult than in the linear J-matrix method wherein the $N \times N$ matrix representation of the total Hamiltonian is energy independent but here $H = H_0 + g\omega^2(E)\Lambda$. A tedious job, which could be done as a first attempt, is to solve Eq. (23) repeatedly at some energy points in a given range. The challenge is to find a procedure to solve Eq. (23) once and for all energies and thus eliminate the second of the two problems indicated below Eq. (3). In the following, we suggest a resolution to this second problem.

For the "inner" function space spanned by $\{\phi_n(x)\}_{n=0}^{N-1}$, we make the basis transformation $\phi_n(x) \mapsto \sum_{m=0}^{N-1} \Omega_{n,m}\phi_m(x)$ where $\Omega$ is an $N \times N$ transformation matrix to be determined. This is equivalent to $f_n(E) \mapsto \sum_{m=0}^{N-1} \Omega_{m,n} f_m(E)$, which in matrix notation reads: $|f\rangle \mapsto \Omega^T |f\rangle$. Equivalently, the matrices $H_0$ and $I$ in Eq. (23) transform as $H_0 \mapsto H_0\Omega^T$ and $I \mapsto I\Omega^T$. Multiplying the equation from left by $\Omega$, they become $\Omega H_0 \Omega^T$ and $\Omega\Omega^T$, respectively. On the other hand, the nonlinear term in the matrix of Eq. (23) is originally a cubic product of $|f\rangle$. Therefore, the matrix $\Lambda$ transforms (after multiplication from left by $\Omega$) to the matrix $(\Omega\Lambda\Omega^T)\Omega\Omega^T$. Now, *we require* that $\Omega\Lambda\Omega^T = I$ making the transformed equation (23) read as follows

$$\left(\begin{array}{c} \\ \\ \\ \hat{H}_0 + g\omega^2(E)\Sigma - E\,\Sigma \\ \\ \\ \\ \end{array}\right)\begin{pmatrix} f_0 \\ f_1 \\ f_2 \\ \times \\ \times \\ f_{N-2} \\ F_{N-1} \end{pmatrix} = \begin{pmatrix} 0 \\ 0 \\ 0 \\ \times \\ \times \\ 0 \\ -b_{N-1}F_N \end{pmatrix}, \qquad (24a)$$

where $\left(\hat{H}_0\right)_{n,m} = \left(\Omega H_0 \Omega^T\right)_{n,m}$ and $\Sigma = \Omega\Omega^T$. If we define the energy variable $\hat{E} := E - g\omega^2(E)$, then Eq. (24a) becomes

$$\begin{pmatrix} & & & \\ & & & \\ & \hat{H}_0 - \hat{E}\,\Sigma & & \\ & & & \\ & & & \end{pmatrix} \begin{pmatrix} f_0 \\ f_1 \\ f_2 \\ \times \\ \times \\ f_{N-2} \\ F_{N-1} \end{pmatrix} = \begin{pmatrix} 0 \\ 0 \\ 0 \\ \times \\ \times \\ 0 \\ -b_{N-1}F_N \end{pmatrix}, \tag{24b}$$

which could be diabolized and solved for $\{f_n(E)\}_{n=0}^{N-2}$ and $\delta(E)$ for all energies $\hat{E}$ using the finite Green's function method (see, for example, Appendix C in Ref. [6]), which goes briefly as follows. If we call the generalized eigenvalues and corresponding normalized eigenvectors of the $N \times N$ matrix pair $\left(\hat{H}_0, \Sigma\right)$ as $\{\varepsilon_n\}_{n=0}^{N-1}$ and $\{\Gamma_{m,n}\}_{m=0}^{N-1}$, respectively, then

$$G_{N-1,N-1}(E) = \sum_{m=0}^{N-1} \frac{1}{\tau_m} \frac{\Gamma_{N-1,m}^2}{\varepsilon_m - \hat{E}}, \tag{25}$$

where $\tau_n = \left(\Gamma^T \Sigma\, \Gamma\right)_{n,n}$. An alternative expression that does not require the calculation of eigenvectors but only eigenvalues, which is numerically preferred, reads as follows

$$G_{N-1,N-1}(E) = \frac{|\tilde{\Sigma}|}{|\Sigma|} \frac{\prod_{m=0}^{N-2} \tilde{\varepsilon}_m - \hat{E}}{\prod_{n=0}^{N-1} \varepsilon_n - \hat{E}} = \left( \frac{\prod_{m=0}^{N-2} \tilde{\xi}_m}{\prod_{n=0}^{N-1} \xi_n} \right) \frac{\prod_{m=0}^{N-2} \tilde{\varepsilon}_m - \hat{E}}{\prod_{n=0}^{N-1} \varepsilon_n - \hat{E}}, \tag{26}$$

where $\{\tilde{\varepsilon}_m\}_{m=0}^{N-2}$ are the generalized eigenvalues of the $(N-1)\times(N-1)$ matrix pair $\left(\hat{H}_0, \Sigma\right)$ after deleting row $(N-1)$ and column $(N-1)$ from both matrices. Moreover, $|\Sigma|$ is the determinant of $\Sigma$ whereas $|\tilde{\Sigma}|$ is the determinant of the same matrix after deleting row $(N-1)$ and column $(N-1)$. On the other hand, $\{\xi_n\}_{n=0}^{N-1}$ are the eigenvalues of the matrix $\Sigma$ whereas $\{\tilde{\xi}_m\}_{m=0}^{N-2}$ are the eigenvalues of the matrix $\tilde{\Sigma}$. To compute the transformation matrix $\Omega$ that satisfies $\Omega\Lambda\Omega^T = I$, we proceed as follows. If we let $U$ be the $N \times N$ matrix whose columns are the normalized eigenvectors of $\Lambda$ then $U^T \Lambda U$ is a diagonal $N \times N$ matrix whose diagonal elements are the *positive* eigenvalues of $\Lambda$ (call them $\{\sigma_n\}$). Consequently, we can write the basis transformation matrix as $\Omega = QU^T$ where $Q$ is a diagonal $N \times N$ matrix whose elements are $Q_{n,m} = \delta_{n,m} / \sqrt{\sigma_n}$ then $\Omega\Lambda\Omega^T = I$.

Substituting $G_{N-1,N-1}(E)$ from (25) or (26) into (15) gives the sought after scattering matrix. For a given physical parameters $\ell$ and $g$, Figure 1 is a plot of $\left|1 - e^{2i\delta(E)}\right|$ with $\omega(E) = \mu^{2\nu} e^{-\mu^2}$ and for several values of the parameter $\nu$. We set the integer $K = 8$ in Eq. (3) and Eq. (17) and took the basis size $N = 20$. The figure indicates the presence of a resonance activity in the energy range $E \approx [3.0, 3.5]$ (a.u.).

# 4. Conclusion and discussion

The J-matrix is an algebraic method for quantum scattering developed in the mid 1970's to handle linear interactions with short range potentials. Its accuracy and convergence properties compares favorably with other successful scattering methods. The J-matrix found a large number of applications in atomic, molecular and nuclear physics. It was extended to multi-channel as well as relativistic scattering, but also in the linear domain. This work constitutes the first attempt at extending the method to handle nonlinear interactions. For simplicity, this development was limited to a simple model where we derived the corresponding scattering matrix. The development relied predominantly on the linearization of products of orthogonal polynomials. The toy model introduced in this work is characterized by three ingredients:

(1) The linear potential $V$ was set to zero.

(2) The assumption that $\alpha_{nm}^{ij} = g\,\delta_{nm}\delta^{ij}$, and

(3) The ansatz for the expansion coefficients $\{f_n(E)\}_{n=0}^{N-2}$ that was proposed in Eq. (16).

This could be generalized and made into a highly non-trivial model. For example, we can add a linear potential $V$ such that the matrix representation of the linear Hamiltonian $H_0 + V$ in the basis $\{\phi_n(x)\}$ is tridiagonal and symmetric. That is,

$$\langle \phi_n | (H_0 + V) | \phi_m \rangle = a_n \delta_{nm} + b_n \delta_{n,m-1} + b_{n-1}\delta_{n,m+1}\,. \tag{27}$$

Moreover, we could take $\alpha_{nm}^{ij\ldots k} = w_{nm}\delta^{ij\ldots k}$ where $w$ is an $N \times N$ real symmetric matrix and $\delta^{ij\ldots k}$ is an $l$-dimensional Kronecker delta corresponding to a nonlinear potential of the form $|\chi(t,x)|^{l+2}$ making $W(E) = w\sum_{i=0}^{K-1}[f_i(E)]^l$. Additionally, we could propose the following generalized ansatz for $\{f_n(E)\}_{n=0}^{N-2}$ to replace Eq. (16)

$$f_n(E) := \sqrt{\left(dz/dE\right)\rho(z)}\left[A_n\, p_n(z)\right], \tag{28}$$

where $z$ is an energy dependent parameter such that $\left(dz/dE\right) > 0$ for $E > 0$ and $\{p_n(z)\}$ is a complete set of orthogonal polynomials with weight function $\rho(z)$. We plan to follow the present work with another that incorporates such generalization.

Finally, we point out a technical difficulty that we faced while calculating the Green's function (25) or (26). For a larger matrix size $N$ and/or a large number of terms $K$ in the sum $\Lambda_{n,m} = \sum_{i=0}^{K-1} D_{i,n}^m(\nu)$, our calculation software (Mathcad® 14.0) fails to produce real eigenvalues $\{\varepsilon_n\}$ and eigenvectors $\{\Gamma_{m,n}\}$ as it should. It is likely that another programming strategy that is better than ours or a more robust computational software with higher calculation precision can overcome this technical difficulty.

## Appendix: Proof that the matrix $\Lambda$ is positive definite

In this Appendix, we prove that the matrix $\Lambda$, whose elements are written as the sum $\Lambda_{n,m} = \sum_{i=0}^{K-1} D_{i,n}^{m}(\nu)$ where $D_{i,n}^{m}(\nu)$ is defined in Eq. (19), is a positive definite matrix. That is, all eigenvalues of $\Lambda$ are real and positive. The proof is based on the fact that the tridiagonal symmetric matrix $J$ shown in (20) is non-singular (i.e., its determinant does not vanish or that it is invertible) and Hermitian as long as $\nu > -1$. In fact, one can show that if $\nu > -1$, then $J$ itself is a positive definite matrix.

**Proof**: Let $A$ be an $n \times n$ non-singular square matrix and let $\gamma$ be any one of its $n$ eigenvalues. Then, the corresponding eigenvalue of $A^{\dagger}$, which is the Hermitian conjugate of $A$, is $\gamma^{*}$. Therefore, $A^{\dagger}A$ and $AA^{\dagger}$ will have the non-negative eigenvalue $|\gamma|^2$. In fact, $|\gamma|^2$ is strictly positive (i.e., cannot be zero) since $A$ is non-singular. Accordingly, the matrices $A^{\dagger}A$ and $AA^{\dagger}$ are positive definite. Using Eq. (19), we can write $\Lambda = \sum_{i=0}^{K-1} B_i^2$ where $B_i = \tilde{L}_i^{\nu}(J)$, which is a polynomial in $J$ of degree $i$ with real coefficients. Now, since $J$ is a non-singular Hermitian matrix for $\nu > -1$ then so too is $B_i$. Thus, if we take $A = B_i$ then $A^{\dagger}A = AA^{\dagger} = B_i^2$ is a positive definite matrix for all $i$. Consequently, $\Lambda$ is positive definite since it is the sum of $K$ positive definite matrices. ■

## References

[1] A. D. Alhaidari, *Representation of the quantum mechanical wavefunction by orthogonal polynomials in the energy and physical parameters*, Commun. Theor. Phys. **72** (2020) 015104

[2] A. D. Alhaidari, E. J. Heller, H. A. Yamani, and M. S. Abdelmonem (Eds.), *The J-Matrix Method: Developments and Applications* (Springer, Dordrecht, Netherlands, 2008)

[3] See Section 9 in: M. E. H. Ismail, *Classical and Quantum Orthogonal Polynomials in One Variable* (Cambridge University Press, 2005)

[4] H. A. Yamani and L. Fishman, *J-matrix Method: Extension to Arbitrary Angular Momentum and to Coulomb Scattering*, J. Math. Phys. **16** (1975) 410

[5] A. D. Alhaidari, *Exact and simple formulas for the linearization coefficients of products of orthogonal polynomials and physical application*, submitted

[6] A. D. Alhaidari, H. Bahlouli, C. P. Aparicio, and S. M. Al-Marzoug, *J-matrix method of scattering for inverse-square singular potentials with supercritical coupling* I. *No regularization*, Ann. Phys. **444** (2022) 169020

# Figure Caption

**Fig. 1**: Plot of the scattering amplitude $|1-S(E)|$ as function of the energy (in atomic units) for the toy model given by $W_{nm} = g\,\delta_{nm}\sum_{i=0}^{K-1} f_i^2$ with the ansatz (16) where $\omega(E) = \mu^{2\nu} e^{-\mu^2}$ and for several values of the parameter $\nu$. We took the physical parameters $\ell = 1$, $g = 2$ and the computational parameters $N = 20$, $K = 8$ and $\lambda = 5$ (in atomic units). The figure indicates a resonance activity at $E \approx 3.6$ for $\nu = 1$ that shifts to lower energy as $\nu$ increases with $E \approx 3.0$ for $\nu = 7$.

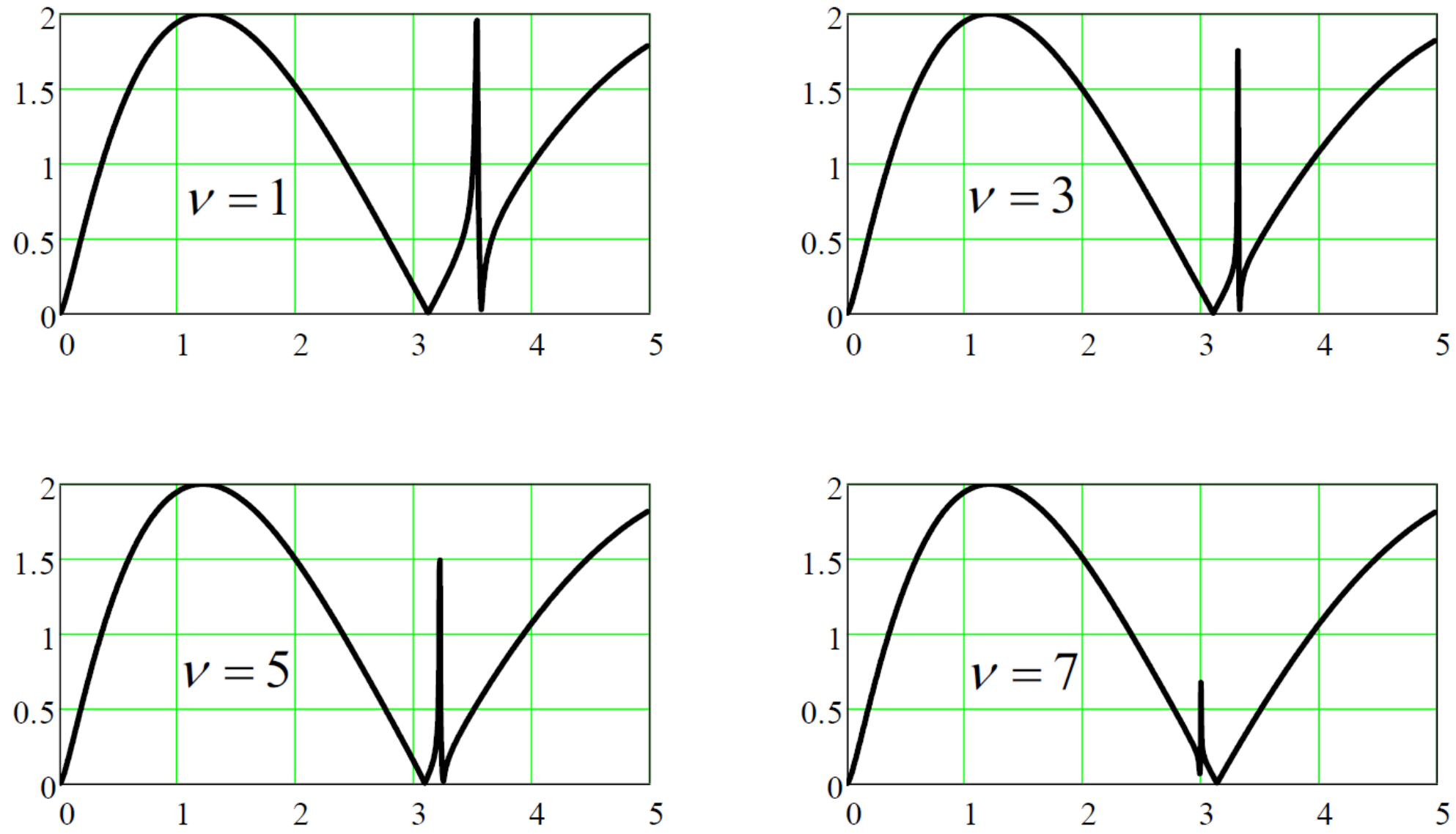

**Fig. 1**